# SSBI-Free Direct Detection via Phase Diverse of Residual Optical Carrier Enabled by Finite Extinction Ratio IQ Modulator for Datacenter Interconnections

XIAOBO ZENG,[1,][*] LIANGCAI CHEN,[1] PAN LIU,[1] AND RUONAN DENG[2]

[1]*Hunan Province Key Laboratory of Credible Intelligent Navigation and Positioning, Key Laboratory of Intelligent Computing and Information Processing of Ministry of Education, National Center for Applied Mathematics in Hunan, Xiangtan University, Xiangtan, 411105, China*
[2]*College of Meteorology and Oceanology, National University of Defense Technology, Changsha, 410073, China*
*\*xiaobo.zeng@xtu.edu.cn*

**Abstract:** Cost-effective, low-complexity and spectrally efficient interconnection can offer fundamental guiding law for future datacenter. In this work, we demonstrate a cost-efficient SSBI-free direct detection for datacenter interconnection, leveraging the phase diversity of residual optical carrier caused by finite-extinction ratio (ER) IQ modulators, combining the device cost-effective IQ modulator with finite-ER and efficient SSBI-free phase-diverse direct detection receiver. Specifically, the proposed solution transforms the inherent limitation of finite-ER of cost-effective IQ modulator into the residual optical carrier advantage of SSBI-free direct detection systems, eliminating SSBI without additional hardware and control complexity. A digital pre-distortion and offset correction algorithms, and a PD-thermal-noise constrained SSBI-free direct detection and signal recovery algorithms are derived and implemented. Comprehensive simulations are conducted. A Global-SNR gain of 1.78 dB and 400 Gb/s data rate are achieved in 100-km SSMF transmission when ($ER_i$, $ER_o$)= (7 dB, 25 dB) of IQ modulator. The proposed solution enables low-complexity, cost-effective, and spectrally-efficient interconnects for next-generation datacenters.

## 1. Introduction

Demand for higher throughput and lower cost per bit drives next-generation datacenter interconnects [1, 2]. Coherent detection, leveraging a local oscillator (LO), enables linear optical-to-electrical field mapping, facilitating chromatic dispersion (CD) compensation via digital signal processing (DSP) [3-5]. To eliminate the LO and reduce device cost, self-coherent receivers, including Kramers-Kronig receiver [6], Stokes-vector receiver [7], carrier-assisted differential detection [8], phase-shifted receiver [9] and phase-diverse receiver [10] are proposed. However, an additional optical carrier or pilot-tone is required to retrieve signal phase or mitigate the signal-signal beating interference (SSBI) caused by the photodiode (PD), which increases the costs of device, power or DSP complexity. For high symbol-rate, and high spectral-efficiency transmissions, transceiver impairments, particularly from low-cost components, significantly degrade system performance [11]. Consequently, receiver design should comprehensively account for impairments of transmitter, including finite extinction ratio (ER) and nonlinearity of in-phase and quadrature (IQ) modulator, to achieve robust performance.

In this paper, we propose and demonstrate a cost-efficient SSBI-free direct detection for datacenter interconnection, leveraging the phase diversity of residual optical carrier caused by finite-ER IQ modulators, which combines the cost-effective IQ modulator and efficient SSBI-free phase-diverse direct detection receiver. The proposed solution transforms the inherent finite-ER limitation of cost-effective IQ modulators into an SSBI-suppression resource, eliminating dedicated optical carrier insertion modules and complex controls. Key contributions include: (1) Auxiliary-carrier-free operation via repurposing intrinsic residual optical carrier, reducing transceiver cost; (2) Theoretical derivation of digital pre-distortion (DPD) and offset correction algorithm, achieving a Global-SNR gain of 1.78 dB over conventional systems with the ER parameters of IQ modulator of $(ER_i, ER_o)$ =(7 dB, 25 dB); (3) Theoretical derivation of the PD-thermal-noise constrained SSBI-free direct detection and signal recovery algorithm, obtaining and verifying optimal phase of $\theta=2\pi/3$, and achieving a high tolerance for the phase deviation. Numerical validation demonstrates the data rate of 400 Gb/s for 100 km standard single-mode fiber (SSMF).

The rest of the paper is organized as follows. In Section 2, the derivations of DPD and offset correction for IQ modulator with finite extinction ratio are thoroughly introduced, which is the basis for SSBI-free direct detection via phase diverse of residual optical carrier. In Section 3, the derivation of noise constrained SSBI-Free direct detection is presented. Section 4 presents the simulation verification, comparing the performance against coherent system. Section 5 concludes this paper.

## 2. Digital pre-distortion for IQ modulator with finite extinction ratio

Fig. 1(a) illustrates the internal structure of IQ modulator, comprising two inner Mach-Zehnder interferometers (MZIs) that modulate the $I$ and $Q$ components. A phase shift of $\pi$ between two arms of each MZI induces destructive interference, achieving near-zero output extinction under ideal conditions. The outer MZI combines $I$ and $Q$ components with a relative phase difference of $\pi/2$, as identified in Fig. 2(b) for the null point of the inner MZI. However, due to the fabrication tolerance, a 50/50 splitting ratio of "Y" splitter and the balanced losses of two arms in an MZI are difficult to achieve, thus resulting in a finite-ER which is defined as

$$ER = ((1+g_m)/(1-g_m))^2, m \in \{I, Q, P\} \quad (1)$$

where $g_{I/Q}$ and $g_P$ characterize the split-ratio and path-loss imbalances of the inner and outer MZIs, respectively. The output of IQ modulator can be expressed as [12]:

$$E_{out}(t) = \sqrt{\frac{g_P}{1+g_P}} e^{j\pi\frac{-V_P}{2V_{\pi,P}}} \left( \sqrt{\frac{g_I}{1+g_I}} e^{j\pi\frac{V_I(t)}{2V_{\pi,I}}} + \sqrt{\frac{1}{1+g_I}} e^{j\pi\frac{-V_I(t)}{2V_{\pi,I}}} \right) +$$
$$\sqrt{\frac{1}{1+g_P}} e^{j\pi\frac{V_P}{2V_{\pi,P}}} \left( \sqrt{\frac{g_Q}{1+g_Q}} e^{j\pi\frac{V_Q(t)}{2V_{\pi,Q}}} + \sqrt{\frac{1}{1+g_Q}} e^{j\pi\frac{-V_Q(t)}{2V_{\pi,Q}}} \right) \quad (2)$$

where $V_{\pi,I}$, $V_{\pi,Q}$ and $V_{\pi,P}$ are the half-wave voltages for the $I$, $Q$ and $P$ MZI, respectively. $V_I(t)$, $V_Q(t)$ and $V_P(t)$ are the driving signals applied to $I$, $Q$ and $P$ MZI, respectively, including the bias voltage. For an ideal MZM, the $ER \to \infty$ and $g_{I/Q/P} = 1$, enabling perfect residual optical carrier suppression, as shown in Fig. 2(c) with $ER$ = inf. However, fabrication-induced variations limit practical $ER$ of MZI to a finite value. Consequently, a residual optical carrier is generated along with the modulated optical signal, as shown in Fig. 2(c) with the $(ER_i, ER_o)$(dB) =($\infty, \infty$), $(10, \infty)$, $(\infty, 10)$, $(10, 10)$, respectively, where the $ER_i$ and $ER_o$ represent $ER_{inner}$ and $ER_{outer}$ of the inner-MZI and outer-MZI, respectively. In this work, a closed-form DPD and offset correction algorithm are theoretically derived, and used to pre-compensate the impairments of nonlinearity and finite-ER of IQ modulator. The pre-distorted signals of $v_I$ and $v_Q$ can be expressed as

$$v_I(t) = (\frac{V_{\pi,I}}{\pi})(\arcsin(\text{sign}(\theta_I) \cdot \min(|\theta_I|, 1)) - \frac{\pi}{2}) - V_{B,I} \tag{3}$$

$$v_Q(t) = (\frac{V_{\pi,Q}}{\pi})(\arcsin(\text{sign}(\theta_Q) \cdot \min(|\theta_Q|, 1)) - \frac{\pi}{2}) - V_{B,Q} \tag{4}$$

where sign(·) denotes the sign function, $V_{B,I}$ and $V_{B,Q}$ are the bias voltages applied to the $I$ and $Q$ MZI, respectively. The signals of $\theta_I$, $\theta_Q$, and the DPD coefficients of $a_I$, $b_I$, $c_I$, $d_I$, $a_Q$, $b_Q$, $c_Q$, and $d_Q$ can be expressed as [12]

$$\theta_I = a_I s_I + b_I s_Q^2 + d_I s_Q + c_I \tag{5}$$

$$\theta_Q = a_Q s_Q + b_Q s_I^2 + d_Q s_I + c_Q \tag{6}$$

where $s_I$ and $s_Q$ are the electrical signals inputted into the $I$ and Q MZI.

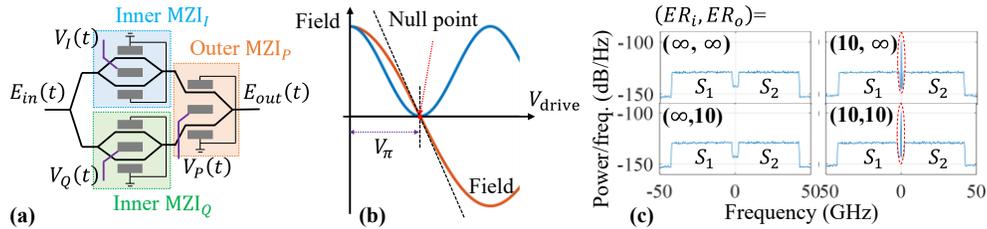

Fig. 1. (a) Structure of IQ modulator in the transmitter. (b) Characteristics of Mach-Zehnder modulator. (c) Example spectra of $(ER_i, ER_o)$=( $\infty, \infty$), ( 10, $\infty$), ( $\infty, 10$), ( 10,10), where $ER_i$ and $ER_o$ represent $ER_{inner}$ and $ER_{outer}$, respectively.

$$a_I = \sqrt{(1+g_P)(1+g_I)}/(\sqrt{g_P} \cdot (1+\sqrt{g_I})) \tag{7}$$

$$b_I = \frac{1}{2} \cdot \frac{\sqrt{(1+g_P)(1+g_I)}}{\sqrt{g_P} \cdot (1+\sqrt{g_I})} \cdot \frac{\sqrt{g_Q}-1}{\sqrt{(1+g_P)(1+g_Q)}} \cdot a_Q^2 \tag{8}$$

$$c_I = -1 \cdot \frac{\sqrt{(1+g_P)(1+g_I)}}{\sqrt{g_P} \cdot (1+\sqrt{g_I})} \cdot \frac{\sqrt{g_Q}-1}{\sqrt{(1+g_P)(1+g_Q)}} \tag{9}$$

$$d_I = \frac{\sqrt{(1+g_P)(1+g_I)}}{\sqrt{g_P} \cdot (1+\sqrt{g_I})} \cdot \frac{\sqrt{g_Q}-1}{\sqrt{(1+g_P)(1+g_Q)}} \cdot a_Q c_Q \tag{10}$$

$$a_Q = \sqrt{(1+g_P)(1+g_Q)}/(1+\sqrt{g_Q}) \tag{11}$$

$$b_Q = \frac{1}{2} \cdot \frac{\sqrt{(1+g_P)(1+g_Q)}}{1+\sqrt{g_Q}} \cdot \frac{\sqrt{g_P} \cdot (1-\sqrt{g_I})}{\sqrt{(1+g_P)(1+g_I)}} \cdot a_I^2 \tag{12}$$

$$c_Q = -1 \cdot \frac{\sqrt{(1+g_P)(1+g_Q)}}{1+\sqrt{g_Q}} \cdot \frac{\sqrt{g_P} \cdot (1-\sqrt{g_I})}{\sqrt{(1+g_P)(1+g_I)}} \tag{13}$$

$$d_Q = \frac{\sqrt{(1+g_P)(1+g_Q)}}{1+\sqrt{g_Q}} \cdot \frac{\sqrt{g_P} \cdot (1-\sqrt{g_I})}{\sqrt{(1+g_P)(1+g_I)}} \cdot a_I c_I \tag{14}$$

For the offset correction algorithm, an offset correction factor of $\alpha$ is used to precisely tune the power of residual optical carrier for achieving an optimal performance. The pre-distorted signals of $v_I$ and $v_Q$ can be updated and expressed as:

$$v_{I/Q,t} = v_{I/Q} - \alpha \cdot \text{sign}(\langle v_{I/Q} \rangle) \cdot \langle v_{I/Q} \rangle \tag{15}$$

where $v_{I/Q}$ is the pre-distorted signals as expressed in (3) and (4). $\langle \cdot \rangle$ represents the mean operation. sign(·) is the signum function.

## 3. Noise constrained SSBI-Free direct detection via phase diverse of residual optical carrier

Fig. 2(a) depicts the structure of residual-optical carrier phase-diverse direct detection receiver, where the input optical signal is split equally into 3 branches. The upper branch implements a narrow-band phase shift of $\theta$ on the central residual-optical carrier, the bottom branch incorporates an equal but opposite phase shift of $-\theta$, while the middle branch retains the unmodified carrier. This configuration yields three optical outputs and are expressed as[10]

$$s_{o1} = Ce^{j\theta} + s(t) \tag{16}$$

$$s_{o2} = C + s(t) \tag{17}$$

$$s_{o3} = Ce^{-j\theta} + s(t) \tag{18}$$

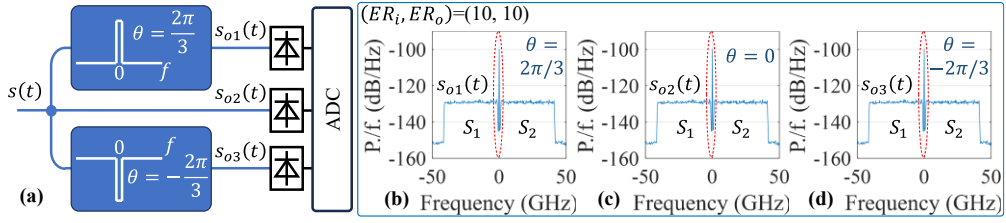

Fig. 2. (a) Structure of phase-diverse direct detection receiver. (b)/(c)/(d) Spectra of $s_{o1}$, $s_{o2}$ and $s_{o3}$ with the phase shifts on the central residual-optical carrier of $2\pi/3$, $0$, $-2\pi/3$, respectively.

where $C$ denotes the residual optical carrier for phase reference. $s(t)$ represents the target information-bearing complex signal. Following the PD detection, photocurrents signal with the thermal noise induced by PD can be derived as

$$i_1 = |s_{o1}|^2 + n_1 = |C|^2 + 2Re\{C^* e^{-j\theta} s(t)\} + |s(t)|^2 + n_1 \tag{19}$$
$$i_2 = |s_{o2}|^2 + n_2 = |C|^2 + 2Re\{C^* s(t)\} + |s(t)|^2 + n_2 \tag{20}$$
$$i_3 = |s_{o3}|^2 + n_3 = |C|^2 + 2Re\{C^* e^{j\theta} s(t)\} + |s(t)|^2 + n_3 \tag{21}$$

where $n_k, k \in \{1,2,3\}$ represents the thermal noise induced by PD, modeled as an independent and identically distributed (i.i.d.) Gaussian process with zero mean and variance $\delta^2$, i.e., $n_k \sim N(0, \delta^2)$. The recovered signals of $\hat{s}$, $\hat{s}_i$ and $\hat{s}_q$ can be respectively expressed as

$$\hat{s} = \hat{s}_i + j\hat{s}_q = (s_i + n_i) + j(s_q + n_q) \tag{22}$$
$$\hat{s}_i = (i_1 + i_3 - 2i_2)/(4C(\cos\theta - 1)) \tag{23}$$
$$\hat{s}_q = (i_1 - i_3)/(4C\sin\theta) \tag{24}$$

where,

$$Var(n_i) = 3\delta^2/(8C^2(\cos\theta - 1)^2) \tag{25}$$
$$Var(n_q) = \delta^2/(8C^2 \sin^2\theta) \tag{26}$$

For a complex signal of $s(t) = s_i(t) + j \cdot s_q(t)$ with power of $P_s$, the $SNR_i$ and $SNR_q$ are derived as

$$SNR_i = \frac{E[s_i^2]}{Var(n_i)} = \frac{P_s/2}{Var(n_i)} = \frac{4P_s C^2 (\cos\theta - 1)^2}{3\delta^2} \tag{27}$$
$$SNR_q = \frac{E[s_q^2]}{Var(n_q)} = \frac{P_s/2}{Var(n_q)} = \frac{4P_s C^2 \sin^2\theta}{\delta^2} \tag{28}$$

To ensure a balanced performance between the in-phase and quadrature components, $SNR_i = SNR_q$ and

$$3\sin^2\theta = (\cos\theta - 1)^2 \rightarrow \theta = 2\pi/3 \tag{29}$$

The phase deviation of $\Delta\theta$ for aforementioned $\theta$, as well as the branch-dependent phase deviations of $\Delta\theta_1$ and $\Delta\theta_2$ in upper and bottom branches, respectively, affect the performance of the receiver, which is demonstrated in Section 4.2.

## 4. Simulation setup and results

### 4.1 Simulation setup

To verify the effectiveness of our proposed interconnection solution, comprehensive numerical simulations are conducted. The simulation setup is shown in Fig. 3(a). At the transmitter side, an optical carrier is simulated by a single-frequency signal with the linewidth of 100 kHz, relative intensity noise (RIN) factor of -150 dBc/Hz and the optical power of 16 dBm. Then it is fed into a single-polarization IQ modulator with the $V_\pi$=4 V. The modulator is biased and worked at the null point, as shown in Fig. 1(b), which is consistent with the conventional coherent transmitter, simultaneously, eliminating extra hardware control complexity. The frequency response measured from our experimental setup, as shown in Fig. 3(d), is incorporated into the transmitter of simulation setup. The fiber link with 100-km standard single-mode fiber is employed to emulate the datacenter interconnections. The attenuation and dispersion of SSMF are considered, and neglecting the Kerr nonlinear. At the receiver side, the received optical power (ROP) is adjusted via a variable optical attenuator (VOA). The received optical signal is fed into the 3-branch phase-diverse direct detection receiver, and then detected by 3 PDs. The responsivity, dark current, and power spectral density of thermal noise of the PDs are defined as 0.8 A/W, $5 \times 10^{-9}$ A and $10 \times 10^{-12}$ A/Hz$^{1/2}$, respectively. Further, based on the datasheet of Keysight M8194A arbitrary waveform generator [13], the digital-to-analog converter (DAC) and analog-to-digital converter (ADC) are modeled as ideal quantizer with an effective number of bits (ENOB) of 6.

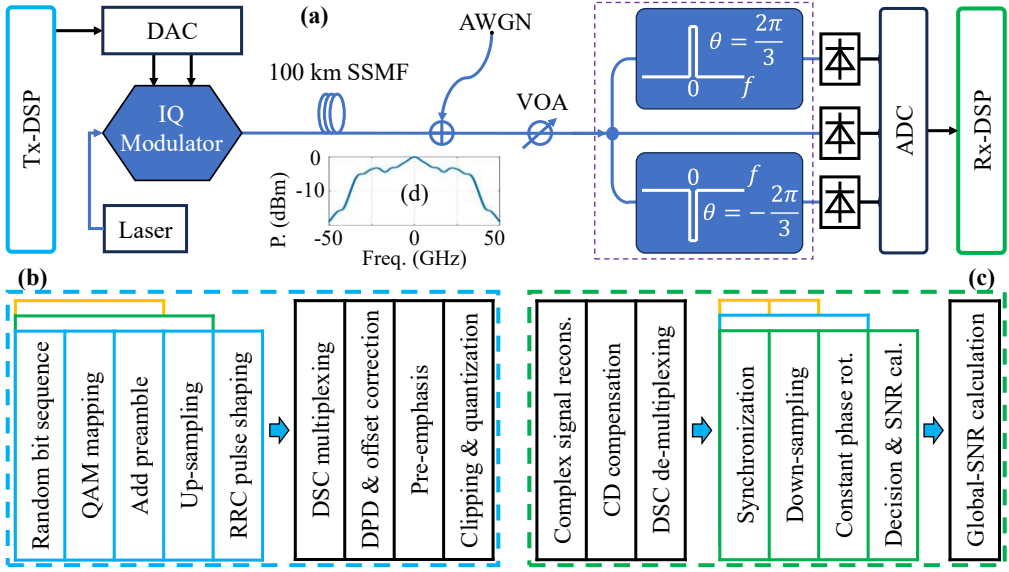

Fig. 3. (a) Simulation setup without electric amplifier, and with the (b) Tx-DSP and (c) Rx-DSP, respectively. (d) Frequency response measured from experimental setup. DSC: digital subcarrier.

For the DSPs at transmitter side (Tx-DSP), as shown in Fig. 3(b), the generated random bit sequence is mapped into symbols with the format of 32-ary quadrature amplitude modulation (QAM) and the length of $2^{16}$, followed by preamble insertion. After up-sampling with 2 samples-per-symbol (sps), the root-raised cosine filter (RRC) based pulse shaping with a roll-off factor of 0.01 is used. In the simulation, 2-band subcarriers with the bandwidth of 40.4 GHz are resampled, and multiplexed with the guard band of 4 GHz. The polynomial DPD and offset correction algorithms described in (2)-(15) are used to pre-compensate the nonlinear distortion, finite-ER degradations, and to correct the signal offsets. The digital pre-emphasis based on zero-forcing is adopted to pre-compensate the bandwidth limitations of transmitter. After clipping, the signals are fed into the DAC and quantized by a quantizer, where the clipping ratio is optimized. For the DSPs at receiver side (Rx-DSP), as shown in Fig. 3(c). After photonic-electric conversion, the complex-value signal $\hat{s}$ can be reconstructed by following the descriptions of (16)-(29). The dispersion compensation is conducted, which can be replaced by a $3 \times 1$ linear least-mean-square (LMS) equalizer, as shown in Fig. 4, to achieve a similar performance. Then, the compensated signal is demultiplexed for separately processing all subcarriers. After synchronization and down-sampling to 1sps, a constant phase rotation is applied to correct the residual phase errors. The symbol decision and signal-to-noise ratio (SNR) calculation are conducted. To comprehensively evaluate the system performance of the proposed solution, Global-SNR is used as a metric, which can be expressed as [14]

$$\text{Global SNR} = \prod_{n=1}^{N}(1 + SNR_n)^{1/N} - 1 \tag{30}$$

where $N$ denotes the number of subcarriers.

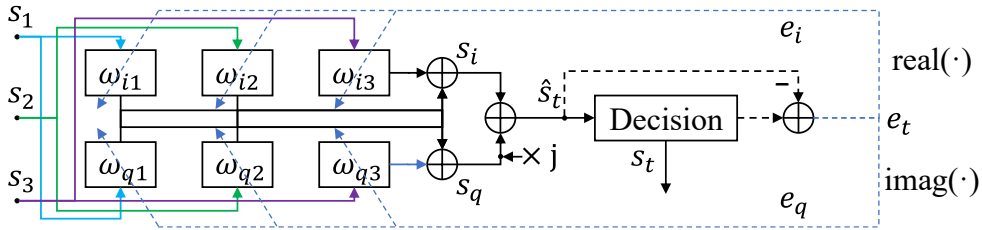

Fig. 4. Structure of 3×1 linear least-mean-square (LMS) equalizer.

### 4.2 Simulation results and analysis

Fig. 5(a) illustrates the recovered Global-SNR versus different scenarios of $(ER_i, ER_o)$ for an IQ modulator and without DPD or offset correction. The $ER_i$ determines the carrier-to-signal power ratio (CSPR), while the $ER_o$ affects the power imbalance between $I$ and $Q$ signals. Consequently, the Global-SNR is dominated by $ER_i$ based on the signal reconstruction algorithm of (16)-(29). The Global-SNR shows a slow increase as $ER_o$= 25 dB for a fixed $ER_i$, while the Global-SNR can achieve the optimal value of 15.89 dB as $ER_i$=7 dB for the fixed $ER_o$ = 25 dB. Fig. 5(b) illustrates the corresponding CSPRs of the modulated signal of $E_{out}$ versus $(ER_i, ER_o)$ of IQ modulator without DPD and offset correction. Due to the nonlinear effects of IQ Modulator, for the optimal Global-SNR at $(ER_i, ER_o)$= (7 dB, 25 dB), the CSPR is -7.34 dB. The spectrum and constellation of the modulated signal are depicted in Fig. 5(e) and (c). The constellation exhibits severe distortion as described in [12], while the constellation of the recovered signal of $S_1$ is illustrated in Fig. 5(d).

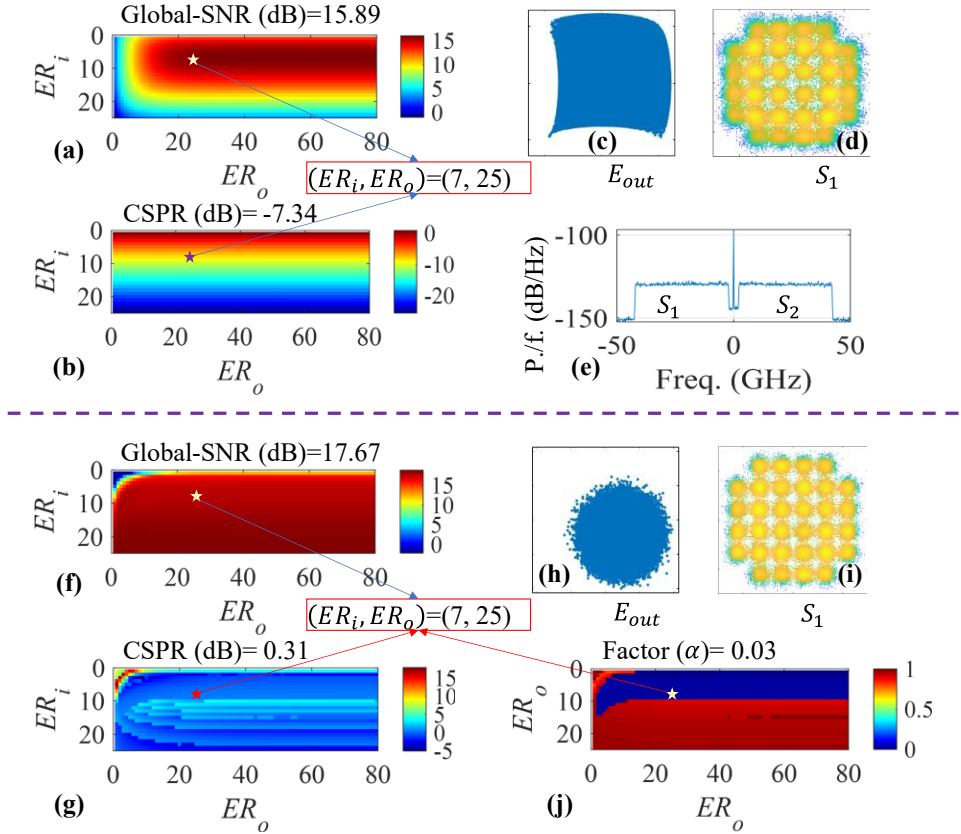

Fig. 5. (a) and (f) Recovered Global-SNR versus $(ER_i, ER_o)$ of IQ modulator without or with DPD and offset correction, respectively. (b) and (g) CSPR versus $(ER_i, ER_o)$ of IQ modulator without or with DPD and offset correction, respectively. (c)/(d) and (h)/(i) Constellations of $E_{out}$ and the recovered signal $S_1$ without or with DPD and offset correction, respectively, when $(ER_i, ER_o) = $ (7 dB, 25 dB). (e) Spectrum of $E_{out}$ when $(ER_i, ER_o) = $ (7 dB, 25 dB). (j) Optimal offset correction factor ($\alpha$) with maximum Global-SNR versus $(ER_i, ER_o)$ of IQ modulator.

To further enhance the Global-SNR and facilitate the regulation of CSPR, the DPD and offset correction algorithms as described in (2)-(15) are adopted. As a comparative experiments, Fig. 5(f) illustrates the Global-SNR for different scenarios of $(ER_i, ER_o)$. Due to the contributions of DPD and offset correction, the Global-SNR of 17.67 dB is achieved as $(ER_i, ER_o)$= (7 dB, 25 dB), thus the gain of 1.78 dB (= 17.67-15.89) is obtained. The CSPR is optimized to 0.31, as shown in Fig. 5(g), which is consistent with the descriptions in [10]. For the optimal Global-SNR, the offset correction factor, $\alpha$=0.03, which is aligning with the descriptions of [14]. The performance improvements can be visually presented through the constellations of Fig. 5(h)/(i) against Fig. 5(c)/(d). The proposed scheme, incorporating DPD and offset correction, reduces $ER_i$ and $ER_o$ to 6 dB and 4 dB, respectively, yielding a Global-SNR of 17.09 dB at $(ER_i, ER_o)$= (6 dB, 4 dB). This represents an improvement of 1.06 dB over the uncorrected system, which achieves a Global-SNR of 16.03 dB. The mitigation of ER relaxes the stringent matching requirements for the "Y" power-splitter and the losses of two arms of IQ modulators. Consequently, the proposed solution accommodates large manufacturing tolerances and reduced device-cost, making it well suited for the cost-effective datacenter interconnections.

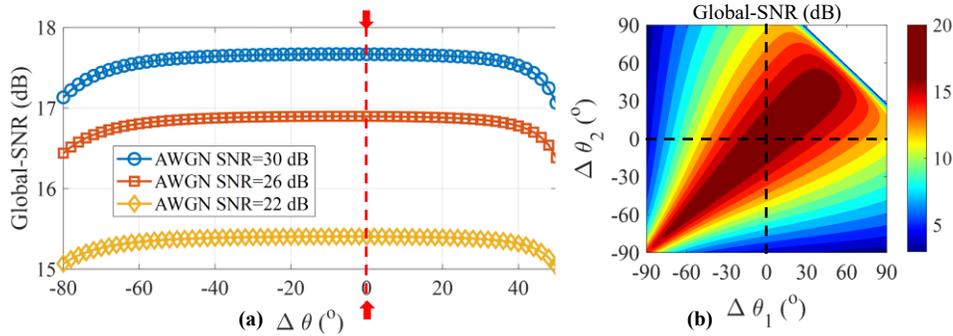

Fig. 6. (a) Global-SNR versus phase deviation $\Delta\theta$(°) for different AWGN noise with the SNRs of 30 dB, 26 dB and 22 dB, respectively, under $(ER_i, ER_o)$= (7 dB, 25 dB) of IQ modulator. (b) Global-SNR as a function of branch-dependent phase deviations of $\Delta\theta_1$ and $\Delta\theta_2$ in upper and bottom branches of receiver under the conditions of AWGN SNR=26 dB and $(ER_i, ER_o)$= (7 dB, 25 dB).

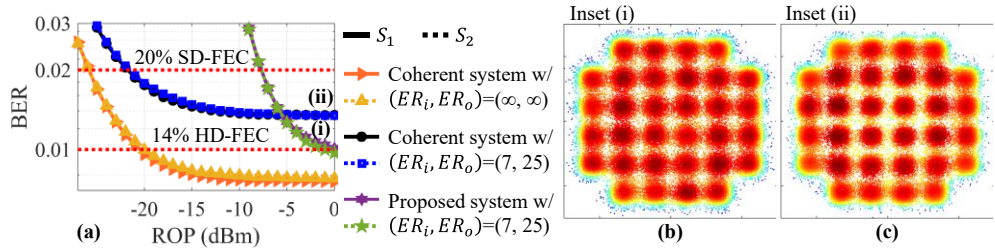

Fig. 7. (a) BER of signals of $S_1$ and $S_2$ versus ROP for fiber length of 100 km when $(ER_i, ER_o)$= (7 dB, 25 dB) and AWGN SNR=26 dB. (b)/(c) Constellations of insets (i) and (ii) for the proposed and coherent system, respectively, with $(ER_i, ER_o)$= (7 dB, 25 dB) and ROP = -1 dBm.

Fig. 6(a) presents the Global-SNR as a function of phase deviation $\Delta\theta(°)$ in the presence of additive white Gaussian noise (AWGN) with SNRs of 30 dB, 26 dB and 22 dB, for an ER pair of $(ER_i, ER_o)$= (7 dB, 25 dB). The maximum Global-SNR occurs at $\Delta\theta = 0°$ and $\theta = 2\pi/3$, in agreement with the theoretical derivation of equations (16)-(29). The simulation results indicate that the proposed solution maintains a Global-SNR above the required threshold over a phase deviation range of $(-60°, 40°)$, demonstrating robust tolerance to phase misalignment of $\Delta\theta$. Furthermore, Fig. 6(b) characterizes the Global-SNR dependence on branch-specific phase deviations of $\Delta\theta_1$ and $\Delta\theta_2$ in the upper and bottom branches of receiver. A symmetric Global-SNR distribution about both axes confirms equal sensitivity to positive and negative phase errors, with the 16.6 dB isocontour confined to $|\Delta\theta_1|=|\Delta\theta_1| < 20^o$. The degradation profile directly correlates with the branch-dependent phase mismatches of receiver, establishing a $\pm 20^o$ tolerance threshold for maintaining Global-SNR > 16.6 dB. This tolerance enables cost-effective manufacturing with minimal calibration requirements for interconnections.

Fig. 7 illustrates the performance comparison between the conventional coherent system and the proposed solution. In Fig. 7(a), the bit-error-rate (BER) of signals $S_1$ and $S_2$ is plotted as a function of ROP for fiber length of 100 km under the conditions of $(ER_i, ER_o)$= (7 dB, 25 dB) and an AWGN SNR of 26 dB. The proposed system achieves the 14% hard-decision forward error correction (HD-FEC) threshold of 1E-2 at an ROP of -1 dBm, and the 20% soft-decision (SD)-FEC of 2E-2 at -7.7 dBm. In contrast, the coherent system employing a finite-extinction-ratio IQ modulator with identical $(ER_i, ER_o)$= (7 dB, 25 dB) fails to reach the HD-FEC threshold, primarily due to residual optical carrier leakage, nonlinear distortions, and I/Q power imbalance introduced by the modulator. Nevertheless, by increasing the complexity of FEC (20% SD-FEC), the non-ideal coherent system can maintain the ROP advantage of 14 dB. Furthermore, the proposed solution supports a data rate of 400 ($= 2 \times 40 \times \log_2 32$) Gb/s. These results demonstrate that the proposed scheme offers a cost-effective transceiver solution with reduced device complexity, albeit with increased power consumption attributable to the phase-diverse optical carrier reception at the receiver.

## 5. Conclusion

In summary, a SSBI-free direct detection solution via phase diverse of residual optical carrier enabled by finite-ER IQ modulator is proposed and demonstrated for datacenter interconnections, which successfully combines the cost-effective IQ modulator with finite-ER and efficient SSBI-free phase-diverse direct detection receiver. The proposed interconnection solution achieves following key contributions: (1) Transforming the inherent defects of low-cost IQ modulators with finite-ER into the residual optical carrier advantage for the SSBI-free direct detection systems, eliminating external insertion modules of optical carrier source or complex control, and effectively reducing the device cost of optical interconnect systems for next-generation datacenters; (2) Deriving and implementing a and offset correction algorithm, and achieving a Global-SNR gain of 1.78 dB at $(ER_i, ER_o)$=(7 dB, 25 dB); (3) Deriving and implementing PD-thermal-noise-constrained SSBI-free direct detection and signal recovery algorithm, obtaining the optimal phase of $\theta = 2\pi/3$, and verifying the accuracy and high tolerance for the phase deviation. The data rate of 400 Gb/s for 100 km SSMF is demonstrated by numerical simulation. The proposed scheme offers a low-complexity, cost-effective, and spectrally efficient interconnect solution for next-generation datacenters.

**Funding.** Project Supported by Provincial Natural Science Foundation of Hunan (2026JJ60507); Project Supported by Scientific Research Fund of Hunan Provincial Education Department (25C0062); Doctoral Research Startup Foundation of Xiangtan University (24QDZ32).

**Acknowledgment.** This work was supported in part by the High-Performance Computing Platform of Xiangtan University.

**Disclosures.** The authors declare no conflicts of interest.

**Data availability.** Data underlying the results presented in this paper are not publicly available at this time but may be obtained from the authors upon reasonable request.